
%
\mag 1200
\vsize=9.6truein \hsize=6.3truein
\parskip =2ex plus .5ex minus .1ex
\parindent 3em

\font\bfnew= cmbx7 at 10pt
\font\big = cmbx9 scaled \magstep1

\textfont\bffam=\bfnew
\def\bf{\fam\bffam\bfnew}
\def\title#1{\vglue -1.0truecm
\centerline{\it Invited review at the STScI meeting on Extragalactic
Background Radiation -- May 1993}
\vglue 2 truecm
\noindent{\big #1}}
\def\author#1{\vglue 1.0truecm\hangindent 2.5truecm
\noindent\hglue 2.5truecm #1\vskip 2.5truecm}
\def\abstract#1{{\narrower\noindent{\bf Abstract.} #1\vskip 2.0truecm}}
\outer\def\sec#1\par{
  \bigbreak
  \vskip 14pt plus2pt minus2pt
  \vskip\parskip\message{#1}\leftline{\big#1}\nobreak
  \smallskip\noindent}
%
%
\def\japref{\parskip=0pt\par\noindent\hangindent1truecm
    \parskip =2ex plus .5ex minus .1ex}
\def\eol{\hfill\break}
\def\gs{\mathrel{\raise1.16pt\hbox{$>$}\kern-7.0pt
\lower3.06pt\hbox{{$\scriptstyle \sim$}}}}
\def\ls{\mathrel{\raise1.16pt\hbox{$<$}\kern-7.0pt
\lower3.06pt\hbox{{$\scriptstyle \sim$}}}}
 at 14.4truept
\def\ss{\rm\scriptscriptstyle}
\font\japrmten = cmr10 at 10.0truept
\def\japfig#1#2#3{\midinsert\vbox{
 \parindent 1truecm
 \vglue#3 truecm
 \narrower
 \lineskiplimit=-10pt
 \baselineskip = 0.39truecm\noindent{\bf Figure #1.}
 \quad {\japrmten #2} }\endinsert
 \medskip}

\title{THE RADIO BACKGROUND: RADIO-LOUD GALAXIES \eol
       AT HIGH AND LOW REDSHIFTS}

\author{J.~A. Peacock\eol
Royal Observatory\eol
Blackford Hill\eol
Edinburgh EH9 3HJ\eol
UK}

\abstract{This paper is in two unequal halves. After
dealing with the possibility of a genuine continuum
background at $\lambda\gs 1$ cm, and showing that
it is unlikely to arise in interesting circumstances,
the remainder of the discussion is devoted to discrete
radio sources, and their consequences for cosmology.
Three main issues are considered: (i) what makes
a galaxy radio loud?; (ii) what do we know about
how the population of radio-loud galaxies has
changed with epoch?; and (iii) what can observations
of high-redshift radio galaxies tell us about
general questions of galaxy formation and evolution?
The main conclusion is that radio galaxies are
remarkably ordinary massive ellipticals. The
high-redshift examples are generally old and red
and do not make good candidates for primaeval galaxies.}

\sec{1. INTRODUCTION}

The purpose of this review is to see what
facts of cosmological interest can be dredged
from wavelengths of above a few cm. In order
to deal with modern research, rather than
ancient history, it will be necessary to
cheat a little and concentrate on the discrete-source
population, rather than genuine smooth backgrounds
-- a strategy adopted by many other speakers at
this meeting. However, to do duty to the advertised
title, we begin with a few comments about
what a non-discrete background might actually
mean, were it to exist. Following this,
the concentration will be on radio galaxies:
why are they active, and how has the degree of
activity changed with redshift? The final sections
attempt to liberate us from the shackle of
the radio waveband altogether, and to ask what
general conclusions may be drawn about
stellar evolution and galaxy formation
from optical/IR data on high-redshift radio
galaxies.

Notation: the Hubble constant, where
quoted explicitly, is given in the form
$h=H_0/100\;\rm kms^{-1}Mpc^{-1}$.
If not otherwise specified, $\Omega=1$ and $h=0.5$ are
assumed.

\sec{2. THE SMOOTH RADIO BACKGROUND}

Malcolm Longair has described how the Cavendish Laboratory
spent the 1960s practicing human sacrifice in order to
determine the extragalactic radio-source background,
with the following approximate result:
$$
I_\nu\simeq 6000 \; \left({\nu\over 1\rm\; GHz}\right)^{-0.8}\;
\rm Jy\, sr^{-1},
$$
to within an uncertainty of about 20\% in amplitude
and 0.1 in spectral index. This background dominates over
the CMB for $\lambda\gs 1$ m, and
is consistent with the
integrated contribution of discrete sources.

On the other hand, it is also not ruled out that a genuine
continuum background might exist at up to 10\% or so
of the above level. What would this mean if it was really so?
The hope would be to learn something about diffuse
intergalactic gas, and there are two standard emission
mechanisms to which we might appeal: synchrotron
radiation and bremsstrahlung.
The parameters available are the density of the
emitting plasma, parameterised by its contribution to
$\Omega$ (in the case of synchrotron radiation,
the electrons would have an assumed power-law energy distribution),
plus the local value of either the magnetic field, $B$
or temperature $T$ -- both of which should scale as $(1+z)^2$.
The resultant background can then be worked out in
the standard way (see Longair 1978). For synchrotron
radiation, we get
$$
I_\nu\simeq10^{14}\; \Omega h\;\left({B\over\rm nT}\right)^{1.8}
\; \left({\nu\over 1\rm\; GHz}\right)^{-0.8}\;
\rm Jy\, sr^{-1}.
$$
What is a plausible value for the intergalactic magnetic field?
It is worth recalling that magnetic fields are very much a
skeleton in the closet of cosmology, since we cannot
easily rule out rather large values -- which would
significantly change our ideas about structure formation,
for example. A nice review of the issue is given by Coles (1992);
he argues that $B$ could be as large as $10^{-4}$ nT. This would
allow observed magnetic fields in astrophysical sources
to be made via compression, rather than dynamo effects,
and would greatly alter the progress of galaxy clustering.
For such a field, the observed background would be produced
with $\Omega h\sim 10^{-3}$. This is an implausibly high
density for a plasma with fully relativistic electrons,
but it is perhaps surprising that the effect is this
close to being interesting.

Turning to bremsstrahlung, one can simply try scaling
old solutions for the X-ray background in which
a `low'-energy flux of around $10^{-3}\;\rm Jy\,sr^{-1}$
is produced by models with $T\simeq 10^8$ K and
$\Omega h^2\simeq 0.1$. Since bremsstrahlung emissivity
scales as $T^{-1/2}$, this implies
$$
I_\nu \simeq 10^3
\; \left(\Omega h^2\right)^2
\; \left({T\over 1\rm K}\right)^{-1/2}
\;\rm Jy\, sr^{-1}.
$$
If we ignore the difficulty in keeping plasma
at such temperatures ionized, this seems the closest
that the radio background is likely to get to
setting constraints on Cold Dark Matter\dots

\sec{3. WHAT MAKES A RADIO-LOUD GALAXY?}

Turning now to the infinitely more interesting issue of the
population of discrete radio sources, we first review
what is known about the causes that lead to
enhanced radio emission.

For orientation, it is convenient to give a sketch of
the population, ordered according to output.
Define $P\equiv \log_{10}L_{21\;\rm cm}/\rm WHz^{-1}sr^{-1}$;
at $P\gs 24$ we find the classical radio galaxies and
quasars, conventionally divided roughly into Fanaroff-Riley (1974)
FRII objects like Cygnus A and compact sources
such as 3C273.
At intermediate powers $23\ls P\ls 24$, we find FRI sources:
twin-jet objects such as 3C31, often lurking in clusters.
At $P\ls 24$, we find all the rest of astronomy:
`radio-quiet' quasars, starburst galaxies and normal galaxies.
We shall be concerned here with the bona-fide radio galaxies
having $P\gs 23$.

Two outstanding systematics of such galaxies have long been
known: they are virtually without exception associated
with elliptical/S0 galaxies, and moreover with the massive
members of this class. This strong tendency for the
probability of strong radio activity to increase with
optical luminosity is illustrated in Figure 1.
It is however interesting that this figure
conceals a more complex behaviour noted by Owen \& White (1991).
They showed that the more powerful FRII sources are actually
{\it less\/} likely in the most massive galaxies --
{\it i.e.} the FR transition shifts to higher radio
power at higher optical power. This may indicate an
influence of the local density on the ability of a
radio jet to remain stable (see Prestage \& Peacock 1989).
Nevertheless, the increased tendency of more
massive galaxies to produce sources of FRI output
or above is not in conflict with this interesting
discovery.

\japfig{1}
{The differential probability distribution of
radio power (for $h=1$), for different bins in
optical luminosity, taken from Sadler, Jenkins
\& Kotanyi (1989). Above $P=$ 21 -- 22, the
probability of radio emission is a very rapidly
rising function of optical luminosity.}
{10.0}

The rather narrow spread in stellar luminosity for
radio galaxies has long been known, and is perhaps
best illustrated in the infrared Hubble diagram
(Lilly \& Longair 1984), which displays an rms
of only 0.4 mag. The average absolute magnitude is
somewhat brighter than for normal ellipticals.
The most direct way of demonstrating this is not to
rely on local samples, where powerful radio galaxies
are rare, but to turn to a direct comparison at
intermediate redshifts. Arag\'on-Salamanca {\it et al.}
(1991) give $K$-band data on ellipticals in the A370
cluster at $z=0.37$, from which a Schechter $K^*=16.3$
for the ellipticals can be determined. Lilly \& Longair
give $K=15.2$ for the mean radio galaxy at this redshift,
but this is in a $10''$ diameter aperture, whereas
the cluster data are in $4''\!.8$ apertures. The
aperture correction at this radius is well modelled
by $L\propto r^{0.5}$, which introduces a small (0.4 mag.)
correction, and leads to the conclusion
$$
\langle L_{\ss RG} \rangle \simeq 1.9 L^*_{\ss E}.
$$
Both the large size of this mean luminosity and its
small dispersion may be understood quantitatively as
empirical manifestations of the strong trend of radio
activity with optical luminosity. If we say that the
probability of a galaxy hosting a strong radio AGN
is $P\propto L^\beta_{\rm opt}$, then multiplying
this rising power law into the exponential truncation of
a Schechter function for the elliptical population
as a whole gives roughly the observed mean luminosity
and scatter if $\beta\simeq 4$ -- 5. This may
also go a good deal of the way towards explaining
the dominance of elliptical hosts: the $L*$ values for
elliptical galaxies tend to be a few tenths of a
magnitude brighter than for spirals, a gap which may
be stretched to as much as a magnitude if we
allow for typical bulge-to-disk ratios to obtain
the $L^*$ ratios between ellipticals and spiral
bulges. We would then predict $N_{\ss E}/N_{\ss S}
\simeq 2.5^\beta \sim 10^2$. In other words, massive ellipticals
dominate powerful radio sources because only they have the
exceptionally deep potential wells needed for the most
active radio AGN. This is far from the whole story:
first, any possible spiral identifications for
powerful radio sources must be more at the $\ls 10^{-3}$
level; second, the whole reasoning rests on the
strong $L^\beta$ trend which remains unexplained.
There is still a major puzzle here.

Are there any other distinguishing features of
radio galaxies? Almost all other peculiarities
can either be traced directly to the effect
of the AGN (such as the strong narrow emission
lines), or to the peculiarity of high mass
already discussed. It would be important to know if there
were any systematic differences between those galaxies
that turn on a radio active nucleus and those
that do not -- but there is no strong evidence
for any such difference. Various suggestions have
been made, but these have usually turned out to
be small and subtle effects, whose reality
generates controversy.

For example, about a decade ago it was suggested
that radio-loud galaxies were redder by about
0.03 mag. in $B-V$ (Sparks 1983), rounder (Disney
\& Sparks 1984), more rapidly rotating (Jenkins 1984)
and in denser environments (Sparks {\it et al.} 1984)
than their radio-quiet counterparts of the
same optical luminosity.
Sparks {\it et al.} (1984) argued that these
trends could be understood within a single
picture of fuel gathering in potential wells,
with the deeper wells being more successful
at generating radio activity. However, in subsequent
years the picture has become somewhat more
complicated as further data have accumulated.
For example, Heckman {\it et al.}
(1985) found that the suggestion of excess rotation
was due to a few incorrect measurements in the
compilation used by Jenkins. Smith \& Heckman (1989)
found a normal distribution of axial ratios and
claimed that galaxies were bluer -- sometimes by
as much as 0.2 mag. in $B-V$.
Finally, Smith \& Heckman (1990) found
environments consistent with those of radio-quiet ellipticals.
Part of the problem here may be that any peculiarity
may be a function of radio power, so that different
studies can yield different answers unless they
use the same definition of radio-loudness.
Also, the range of properties of radio-quiet
ellipticals is large and diverse; misleading
conclusions may be reached unless there is a
large and complete comparison sample.
What is needed is a large sample of radio-loud
ellipticals whose properties can be compared to a
radio-quiet set matched in optical luminosity
and redshift.

In the meantime, claimed peculiarities of
radio ellipticals need to be treated with caution.
Two properties which are presently in this provisional
class are the suggestion that radio ellipticals
have low-level isophote distortions indicative
of merging (Smith \& Heckman 1989), and the
question of dynamics. Smith, Heckman \& Illingworth (1990)
found that radio-loud ellipticals lie on the
`fundamental plane' in size/luminosity/velocity dispersion
space, but there are some suggestions that they
may occupy a different region of the plane --
being brighter at a given velocity or size
(Sansom, Wall \& Sparks 1987; Romanishin \& Hintzin 1989).
It will be interesting to see how these suggestions
hold up. We certainly badly need some clear set
of clues as to what triggers these objects.

\sec{4. LUMINOSITY FUNCTIONS}

Now consider what we know empirically about the
abundance of radio AGN at high redshift, and
what constraints this information may set on models
of structure formation.

\sec{\bf 4.1 Observational results}

No significant new datasets relevant to the luminosity
function of powerful radio sources have been published
since the study of the RLF published by James Dunlop
and myself in 1990.
This was based on nearly-complete redshift data on
roughly 500 sources down to a limit of 100 mJy at
2.7 GHz, plus fainter number-count data and
partial identification statistics.

The main conclusions of this study were firstly to
affirm long-standing results (Longair 1966; Wall,
Pearson \& Longair 1980)
that the RLF undergoes differential evolution: the highest luminosity
sources change their comoving densities fastest.
Nevertheless, because the RLF curves, the results
can be described by a model of pure luminosity
evolution for the high-power population,
in close analogy with the situation for
optically-selected quasars (Boyle {\it et al.} 1987).
The characteristic luminosity in this case increases
by a factor $\simeq 20$ between the present and a redshift
of 2. Similar behaviour applies for both steep-spectrum
and flat-spectrum sources, which provides some comfort
for those wedded to unified models for the AGN population.
There is a remarkable similarity here to the
evolution of `starburst'
galaxies, distinguished by blue optical-UV continua and
strong emission from dust which make them very bright in the
IRAS 60-$\mu$m band. It has been increasingly clear since the
work of Windhorst (1984) that such galaxies make up a
substantial part of the radio-source population below
$S\simeq 1$ mJy.
The evolution of these objects at radio wavelengths
and at 60 $\mu$m is directly related because
there exists an excellent correlation between output at
these two wavebands.
Rowan-Robinson {\it et al.}
(1993) have exploited this to investigate the implications
of IRAS evolution for the faint radio counts.
They find good consistency with the
luminosity evolution $L\propto (1+z)^3$ reported for
the complete `QDOT' sample of IRAS galaxies
by Saunders {\it et al.} (1990).

Were it not for the fact that some populations of objects
show little evolution ({\it e.g.} normal galaxies in the near-infrared:
Glazebrook 1991), one might be tempted to suggest an
incorrect cosmological model as the source of this
near-universal behaviour. The alternative is to look for
an explanation which owes more to global changes in the Universe
than in the detailed functioning of AGN.
One obvious candidate, long suspected of playing a role in AGN,
is galaxy mergers; Carlberg (1990) suggested that this mechanism
could provide evolution at about the
right rate (although see Lacey \& Cole 1993). Why the
evolution does not look like density evolution is still a
major stumbling block, but it seems that we should be looking
at this area quite intensively, given that mergers have
been implicated in both AGN and starbursts, and that there
may be some evidence for their operation from the general
galaxy population (Broadhurst, Ellis \& Glazebrook 1992).

However, it is unclear how much emphasis
should be placed on this apparent universality;
particularly, limited statistics make it
uncertain just how well luminosity evolution
is obeyed. For example, Goldschmidt {\it et al.}
(1992) have produced evidence that
the PG survey is very seriously incomplete at
$z\ls 1$; if confirmed, this would imply that
the evolution of quasars of the very highest
luminosities is {\it less\/} than for those a
few magnitudes weaker. Furthermore, the QDOT
database was afflicted by an error in which 10\%
of the galaxies were assigned incorrectly
high redshifts (Lawrence, private communication);
this will probably weaken the IRAS degree of evolution.
It may well be that the degree of unanimity
described above will prove spurious, and that we will
be left with the unsurprising situation that
a complex phenomenon like AGN evolution can only
be described simply when the samples are too
small to show much of the detail.

\sec{\bf 4.2 Redshift cutoff and interpretation}

At higher redshifts, the uncertainties increase as the
data thin out, but there is evidence that the
luminosity function cannot stay at its $z=2$ value at
all higher redshifts. The form of this `redshift cutoff'
is uncertain: we cannot at present distinguish between possibilities
such as a gradual decline for $z>2$, or a constant RLF up
to some critical redshift, followed by a more precipitous decline.
We therefore present a `straw man' model designed to
concentrate the minds of observers, in which the
luminosity evolution goes into reverse at $z\simeq 2$
and the characteristic luminosity retreats by a factor $\simeq 3$
by $z=4$ (Figure 2).

\japfig{2}
{The evolving RLF, according to the pure
luminosity evolution model of Dunlop \& Peacock (1990).
The main features are a break which moves to higher powers
at high redshift, but which declines slightly at $z\gs 2$.
the strength of the break and the rate of evolution are
comparable for both radio spectral classes.
}{13.0}

This model predicts the following fraction of
objects at $z>3.5$ as a function of 1.4-GHz flux-density limit:
0.5\% at 100 mJy; 3\% at 1 mJy. Without some form of
cutoff, these numbers would be about a factor of 5 higher.
The reason for the increased ease of detecting a cutoff at low
flux density is that the RLF
is rather flat at low powers; for $\rho\propto P^{-\beta}$ and
$S\propto \nu^{-\alpha}$, we expect $dN/dz\propto (1+z)^
{-\beta(2+\alpha)-1/2}$. Steep spectra and a steep RLF thus
discriminate against high redshifts, but at low powers the
flatter RLF helps us to see whatever high-$z$ objects there
are more easily.
It should be relatively easy to test for the presence
of a cutoff on the basis of these predictions. This is especially
true at low flux densities (see Figure 3). Here, we still sample
the flat portion of the RLF even at high redshift, and so
the predicted numbers of high-redshift sources is large
without a cutoff -- around 15\% at $z>4$ for a sample at 1 mJy.

\japfig{3}
{A plot of the integral redshift distributions predicted
for two samples limited at 1.4-GHz flux densities of
100 mJy and 1 mJy. The upper line shows a prediction
for a luminosity function which is held constant for
$z\gs 2$; the lower line shows the prediction of the
`negative luminosity evolution' model of Dunlop \& Peacock (1990).
}{13.0}

Whether or not the redshift cutoff is real, we seem
to have direct evidence that the characteristic
comoving density of radio galaxies has not altered
greatly between $z\simeq 4$ and the present. Integrating
to 1 power of 10 below the break in the RLF, we find
$$
\rho\simeq 10^{-6}h^3\;{\rm Mpc}^{-3}.
$$
Is this a surprising number? In models involving
hierarchical collapse, the characteristic mass of
bound objects is an increasing function of time. At high
mass, the abundance of objects falls exponentially
if the statistics of the density field are Gaussian.
Clearly, a model such as CDM (which falls in this class)
will be embarrassed if the density of massive objects
stays high to indefinite redshifts. The analysis of
this problem, using the Press-Schechter mass-function
formalism (Press \& Schechter 1974) was first given by Efstathiou
\& Rees (1988) for optically-selected quasars.

There are two degrees of freedom in the analysis: what
mass of object is under study, and what are the parameters
of the fluctuation power spectrum?
For the first, Efstathiou \& Rees had to construct a long
and uncertain chain of inference leading from quasar
energy output, to black-hole mass, to baryonic galaxy mass,
to total halo mass. For radio galaxies, things are much simpler,
because we can see the galaxy directly.
Infrared observations imply that, certainly up to $z=2$,
the stellar mass of radio galaxies has not changed
significantly. At low redshift,
there is direct evidence that the mass of radio galaxies
exceeds $10^{12}\;M_\odot$, so it seems reasonable
to adopt this value at higher redshift.
Figure 4 shows the Press-Schechter predictions for
two COBE-normalized CDM models. The low-$h$ model which
fits the shape of the galaxy-clustering power spectrum
(Peacock 1991) intersects the observed number density
at low-ish redshifts (7 -- 8), whereas the `standard'
$h=0.5$ model with its higher degree of small-scale
power predicts many more objects.
This is clearly only a suggestive coincidence at
present, but it is clearly interesting that the
model which most nearly describes large-scale
structure also predicts that the formation of massive
objects should occur near the point at which we
infer a lack of high-$z$ AGN.

\japfig{4}
{The epoch dependence of the integral mass function in
CDM, calculated using the Press-Schechter formalism as
in Efstathiou \& Rees (1988). The normalization is to the
COBE detection of CMB fluctuations. Results are
shown for two Hubble constants: the `standard'
$\Omega h=0.5$ (upper panel) and $\Omega h=0.3$ (lower panel).
Here, $\Omega h$ is merely a fitting parameter used
to describe the shape of the power spectrum, and it
does not presuppose a true value of the Hubble constant.
The vertical scaling of density with $h$ is given
explicitly, and the mass values assume $h=0.5$.
The extra small-scale power in the
former case means that many more massive hosts than
the observed radio-galaxy number (horizontal line)
are predicted, even at $z\gs 10$.
}{13.0}

\sec{\bf 4.3 Black-Hole abundances}

In the spirit of this meeting, it is probably important
to concentrate on integrated properties of the
radio-source population. One important feature of
this sort is the relic density of black holes
deposited by the work of past AGN. This is
something  which has been discussed extensively for
radio-quiet quasars, but which has not been given so
much attention in the radio waveband alone. The
advantage of doing this is that, as discussed above,
we have a rather good idea of which galaxies host
radio-loud AGN, and therefore we know where to look
for any debris from burned-out AGN.
The basic analysis of this problem goes
back to Soltan (1982). He showed that the relic
black-hole density may be deduced observationally
in a model-independent manner, as follows.

The mass deposited into black holes in time
$dt$ by an AGN of luminosity $L_\nu$ is
$$
d[M_\bullet c^2]=\epsilon^{-1}g\,[\nu L_\nu]\; dt,
$$
where $\epsilon$ is an efficiency, and $g$ is a
bolometric correction.
To obtain the total mass density in black holes,
we have to multiply the above equation by the
luminosity function (which already gives the comoving
density, as required) and integrate over luminosity.
The integral can be converted to one over redshift
and flux density, and the integrand depends of the
observable distribution of redshifts and flux
densities, so the answer is model dependent.
Doing this for the Radio LF gives a much lower answer
than for optically-selected QSOs, which have a much higher
density:
$$
\eqalign{
\rho_\bullet &= 10^{11.7}\, \epsilon^{-1} g\; \rm M_\odot Gpc^{-3}\quad
(QSO)\cr
\rho_\bullet &= \,10^{9.0}\; \epsilon^{-1} g\; \rm M_\odot Gpc^{-3}\quad
(Radio)\cr
}
$$
Since we know rather well the present density of massive elliptical
galaxies ({\it e.g.} Loveday {\it et al.} 1992),
we may distribute half the above radio mass into
ellipticals above the median radio-galaxy luminosity,
with the following result for the mean hole mass:
$$
\left\langle M_\bullet \right\rangle \simeq
2000 \, \epsilon^{-1} g\; h^{-3}\, \rm M_\odot.
$$
What is the bolometric correction for radio
galaxies? We know that the total output generally peaks in the
IRAS wavelength regime, with an effective $g\sim 100$
(Heckman, Chambers \& Postman 1992); this gives
$$
\left\langle M_\bullet \right\rangle \simeq
2\times 10^5 \, \epsilon^{-1} \, h^{-3}\, \rm M_\odot,
$$
which paints a rather less optimistic prospect for detection
than studies based on the output of QSOs. This is because, even
with such a large $g$, the actual energy radiated by radio
galaxies is rather low, and this is not compensated for fully by
the relative rareness of the host galaxies. The
above figure is not easy to reconcile with large black-hole
masses suggested for some radio AGN. For example,
Lauer {\it et al.} (1992) suggest a central mass of
$M_\bullet \simeq 3\times 10^9 \rm \, M_\odot$ for M87.
Without suggesting that M87 is greatly atypical, this
can always be made consistent by assuming a low enough
efficiency. However, this would not fit well with the
view that radio galaxies are powered via electrodynamic
extraction of black-hole rotational energy ({\it e.g.} Blandford 1990);
here the efficiency can be up to $\epsilon=1-2^{-1/2}$.
If masses of order $10^9\rm M_\odot$ are substantiated
in several radio galaxies or radio-quiet massive ellipticals,
this would be quite a puzzle. Probably the simplest
solution would be to suggest that the total energy was
higher than suggested by the above sum -- perhaps
because radio ellipticals spend part of
their lives as QSOs, where the total energy output
would be considerably higher for a given radio power.

\sec{5. HI SEARCHES FOR HIGH-REDSHIFT GALAXIES}

We now turn to the question of what the radio waveband
has to say about the properties of galaxies seen at
high redshifts.
One unique capability of radio astronomy for cosmology is
the detection of neutral hydrogen via the 21 cm line.
This tends to receive most attention at low redshifts
via the Tully-Fisher relation and the studies of
the distance scale and peculiar velocities.
However, it also gives a unique way of detecting
neutral gas at high redshift -- even beyond the limit
of $z\simeq 5$ where quasar absorption-line studies
can probe. Particularly motivated by early `pancake'
theories of galaxy formation in which purely
baryonic models give a supercluster-scale coherence
length to the mass distribution, there have been a
number of attempts over the years to use low-frequency
observations to detect neutral hydrogen at high redshifts
({\it e.g.} Davies, Pedlar \& Mirabel 1978 [$z=3.3$ \& 4.9];
Bebbington 1986 [$z=8.4$]; Uson, Bagri \& Cornwell 1991 [$z=3.3$];
Wieringa, de Bruyn \& Katgert 1992 [$z=3.3$]).
These are sensitive only to rather large structures:
for a Gaussian velocity dispersion $\sigma_v$, the
expected flux density is
$$
{S\over \rm mJy}=19.9 \, \left({M_{\rm HI}\over 10^{14}\, \rm M_\odot}
\right)\, \left({\sigma_v\over 10^3\, \rm km\,s^{-1}} \right)^{-1}\,
{h^2 \over D^2\; (1+z)},
$$
where $D$ is comoving distance divided by $c/H_0$ -- {\it e.g.}
$D=2(1-[1+z]^{-1/2})$ in an $\Omega=1$ model.
Since sensitivities of a few mJy are typically attained,
the experiment is sensitive to masses in the range
$10^{14}$ -- $10^{15}$ M$_\odot$.

Most such experiments have yielded only upper limits, but
the VLA experiment of Uson, Bagri \& Cornwell (1991) claimed
the detection of a resolved object with a peak flux density of 10 mJy.
The inferred parameters of their object were
$$
\eqalign{
&z = 3.397 \cr
&M_{\rm HI} \simeq 10^{14}\, h^{-2}\, \rm M_\odot\cr
&\theta \simeq 5' \simeq 1\; h^{-1}\; \rm proper\ Mpc\cr
&\sigma_v \simeq 77\;\rm km\,s^{-1}.\cr
}
$$
This experiment caused much debate, particularly the authors'
claim that this was an example of a Zeldovich pancake.
The characteristics of the emission are certainly hard
to understand in any other way. The gas mass and size of
object, together with the effective volume of space
surveyed, are about right for a rich cluster of galaxies.
However, in addition to the minute velocity dispersion,
one would also not expect to find intracluster gas in
a neutral state. In hierarchical models, it is continually
shock heated by new infalling clumps of mass as structure
grows. The only neutral gas would be associated with
individual galaxies, producing much less massive
neutral condensations ({\it e.g.} Subrahmanian \& Padmanabhan 1993).
The only possibility might be a group of unusually
neutral-rich galaxies resembling the damped Lyman-$\alpha$
absorption systems seen in quasar spectra; in this context,
it is worth noting that Wolfe (1993) has shown these to
lie in regions of high density (at least in terms of
cross-correlation with weaker Lyman-$\alpha$ emitters).
In any case, it would still be necessary to appeal
to the coincidence of seeing a cluster close
to its turn-round time to explain why the
velocity dispersion is so small (and even this does
not solve things completely, since there will
be a dispersion associated with substructure).

Only in models with an initial coherence length
does the gas have time to cool and regain its neutrality
following heating at the initial collapse of the cluster.
Without attempting to turn history back to a time
before dark matter, perhaps the least radical
modification would involve warm dark matter
with a coherence length of a few Mpc. This would
in any case lead to the usual `top-down' chain of
events for galaxy formation. Since we believe that
objects of cluster mass in fact mainly formed relatively
recently (Lacey \& Cole 1993; see also the contribution
to this volume by S. White), this would have important
implications for the ages of galaxies. For this reason,
it is vital that the Uson {\it et al.} object
be either confirmed or shown not to exist. Van der Kruit
(private communication) suggests that the Westerbork
group have indeed failed to detect it, which may cause
some relief to those distressed by the above
discussion. Whatever the eventual outcome, such observations
will continue with increasing sensitivity and will be
capable of setting interesting constraints on conditions
at high redshift.

\sec{6. STELLAR POPULATIONS AT HIGH REDSHIFT}

{\bf 6.1 The golden age}
\medskip

\noindent
Finally, for a line of argument that turns out to lead
in completely the opposite direction -- {\it i.e.} to
galaxy formation at rather high redshift -- we turn
to the stellar populations in high-redshift radio galaxies.
Most of the 1980s constitute a vanished age of innocence
for the radio cosmologists: at this time, they were the
only ones able to find galaxies at $z\gs 1$ in any sort of
numbers.
A series of investigations established
several interesting properties for these objects, in particular

\item{(i)}The well-defined $K$--$z$ relation for 3CR radio galaxies,
consistent
with purely passive evolution of their stellar populations, and
producing 1 mag. of brightening by $z\simeq 1$
(Lilly \& Longair 1984).

\item{(ii)}The large scatter in the optical-IR
(Lilly \& Longair 1984) and optical ({\it e.g.} Spinrad \& Djorgovski 1987)
colours of 3CR galaxies, which was interpreted
as reflecting the occurrence of bursts of star formation
in otherwise passively evolving objects.

\item{(iii)}Lilly (1989) argued for a two-component model in which a bluer
component
was superimposed onto a rather red underlying galaxy. To reproduce the
spectral energy distribution of the red component (and thus of the reddest
radio galaxies)
required ages greater than 1 Gyr, pointing to
high formation redshifts (Lilly 1989, Dunlop {\it et al.} 1989b, Windhorst,
Koo \& Spinrad 1986), although there was some controversy over
the model dependence of the exact ages (Chambers \& Charlot 1990).

\item{(iv)}Perhaps the high-water mark of this period was the
discovery by Lilly (1988) of 0902+34 at $z=3.4$ (at a time when the
galaxy redshift record was 1.8). The apparently red colours of this
object argued for a large enough age that the bulk of the
stars must have formed at $z\gs 6$ -- an inference of enormous
importance for models of galaxy formation.

However, over the last few years a revisionist tendency has
appeared -- leading to all the above achievements being questioned.
Even at the time, there was some doubt whether we could be sure that
the above behaviour was representative of all galaxies.
Fears of a radio-induced bias appeared well founded with the discovery of what
has become known as the `alignment effect':
the realisation that at large
redshifts ($z \gs 0.8$) the optical
and radio axes of many of the most powerful radio galaxies are
aligned (McCarthy
{\it et al.} 1987; Chambers, Miley \& van Breugel 1987).
Near-IR images of 3CR galaxies
appeared to confirm that the infrared morphologies
of these objects were in general just as peculiar as their
optical morphologies (Chambers, Miley \& Joyce 1988;
Eisenhardt \& Chokshi 1990; Eales \& Rawlings 1990).
These discoveries provide
direct evidence of radio-induced `pollution'
of the UV-optical light of radio galaxies, and this
led some authors to suggest
that these sources are thus useless as probes of galaxy evolution
in general ({\it e.g.} Eisenhardt \& Chokshi 1990).

Furthermore, it has become apparent that Lilly's galaxy
0902+34 does not have the properties initially claimed.
The $K$ flux is rather lower than Lilly's measurement,
and a large fraction of this smaller total is contributed by
the [OII] 3727\AA\ line, which is redshifted into the
$K$ window. The result is that the galaxy in fact looks very
young: nearly flat-spectrum with no evidence for
the presence of an old component. On this basis,
and considering other similar objects at extreme
redshifts, Eales {\it et al.} (1993) have argued that
radio galaxies at $z\gs 2$ are in effect protogalaxies
observed in the process of formation.

Before accepting this remarkable reversal of conventional
wisdom, however, it is worth bearing in mind that the
galaxies under discussion are among the most luminous
few dozen radio AGN in the entire universe (inevitably:
they are the high-redshift members of bright samples
with $S\sim 1$ -- 10 Jy). In order to draw any general
conclusions about galaxy formation, it is necessary to
understand the effect the AGN has on the optical/IR
properties of the galaxy within which it is embedded.

\sec{\bf 6.2 Alignments as a function of power}

What is required is to be able to study the properties of
galaxies with a wide range of radio powers, and this
is what James Dunlop \& I have attempted in some recent
work (Dunlop \& Peacock 1993).
In order to eliminate possible confusion with any
epoch dependence, we worked with a redshift band around $z\simeq 1$.
At this redshift, it is relatively easy to select samples
unbiased by optical selection, and the objects are bright
enough that high-quality data can be obtained.
We considered galaxies from two catalogues: 19 high-power
3CR galaxies; 14 low-power comparison galaxies with
$S_{\rm 2.7\;GHz} > 0.1$ Jy
from the Parkes Selected Regions (PSR) (Downes {\it et al.} 1986; Dunlop
{\it et al.} 1989a). The PSR
galaxies are a factor $\simeq$ 20 less radio luminous than their 3CR
counterparts. Radio luminosity is the only significant
difference between the radio properties of the two samples.

Our principal dataset on these galaxies is deep
infrared images, taken with the
$62\times 58$ pixel InSb array camera IRCAM,
on the 3.9m United Kingdom Infrared
Telescope (UKIRT), with the camera operating in
the 0.62-arcsec/pixel mode.
{}From these images, we investigated the extent of the
the alignment effect at $z\simeq 1$.
To avoid subjective factors, the infrared position
angles were determined automatically by using the
moments of the sky-subtracted flux within some
circular aperture.
We decided to vary the diameter of the aperture
to adapt to the size of the radio source, because
there are virtually no examples of optical or IR emission
extending beyond the radio lobes.
If the diameter
of the radio source lay between 5 and 8 arcsec, an aperture equal in
diameter to the radio source was used. If the radio source was greater than
8 arcsec in diameter, an 8 arcsec diameter was used (larger
apertures generally contain foreground objects). If the radio source
was smaller than 5 arcsec in diameter, a 5-arcsec diameter was used.

\japfig{5}{Histograms of (IR -- Radio) position angle differences
for the 3CR and PSR samples. The clear difference seen here
is completely robust to different methods for determining
position angles. It is related to the fact that the PSR
galaxies are also rounder, and generally lacking in an extended
aligned component of blue light.}
{7.0}

Figure 5 shows the resulting IR--radio alignment histogram for the
3CR and PSR subsamples.
The infrared alignment effect is extremely obvious in our data for the 3CR
galaxies, which appears
to contrast with the conclusions of Rigler {\it et al.} (1992).
Much of the apparent discrepancy arises from the fact that we have a larger
sample. Position angles for objects in common generally
agree well, but with some exceptions which are due to
different methods of analysis; Rigler {\it et al.} (1992)
sometimes use a large aperture where their position
angle is affected by companion objects.
In contrast to the 3CR sub-sample, there is no evidence of any
significant alignment between the infrared and radio morphologies
of the PSR galaxies. This result is very robust and quite
obvious given the images: the PSR galaxies are rounder, with
generally little sign of the disturbance evident in many
of the 3CR images.

This argues in favour of the two-component model
advanced by Lilly (1989) and Rigler {\it et al.} (1992).
In this, the underlying galaxy is round, but there is
a component of variable amplitude which is elongated
along the radio axis, and it is this which leads
to the alignment. Our data demonstrate that the
strength of this component correlates well with
radio power, as is perhaps not so surprising in
retrospect. Certainly, several models for the
production of this light exist that
predict a correlation with radio power (scattering,
induced star formation, inverse Compton emission
-- see {\it e.g.} Daly 1992 for a review).
We shall not be concerned here with having to plump for
a specific model, but it is worth noting that
evidence is starting to mount in favour of the
explanation in terms of scattering from a hidden
blazar. The main argument in this direction is
the measurement of polarization with E-vector perpendicular
to the radio axis. The first measurements of this
effect gave very low percentage polarizations,
implying that this could not be the dominant
mechanism. However, with better resolution, imaging
polarimetry is now producing polarized fractions
of $\gs 20$\% in the outer parts of strongly
aligned galaxies (Jannuzi \& Elston 1991; Tadhunter
{\it et al.} 1992; Cimatti {\it et al.} 1993).
Given geometrical dilution, it now seems plausible
that the aligned component results from scattering
in at least some objects.

\sec{\bf 6.3 Colours and ages of radio galaxies}

Having seen that the extent of the aligned component scales
so dramatically with radio power, we now look for other
optical/infrared properties which correlate with power.
Given that the aligned component is often bluer than
the nucleus of the galaxy, we should certainly expect to see
some correlation between colour and power.
A useful way of quantifying the degree of UV activity
was introduced by Lilly (1989).
He assumes that the observed spectrum of a radio
galaxy arises from a combination of two distinct components  --
an `old' population with a well-developed 4000${\rm \AA}$ break, and
a `young' flat-spectrum component.
This simple model can be fitted to the observed colours by varying
one parameter. This is $f_{5000}$: the fractional contribution of the
flat-spectrum component to the
galaxy light at a rest wavelength of $5000{\rm \AA}$.
This method can also be used with some success to
estimate the redshift for objects which lack spectroscopy
(see Lilly 1989; Dunlop \& Peacock 1993).
Some of the PSR objects had their redshifts estimated
in exactly this way: the redder objects with low $f_{5000}$
also have low levels of emission-line activity and so are of
course the hardest spectroscopic targets.

\japfig{6}{Two examples of the
spectral fitting used to determine estimated redshifts
and $f_{5000}$, the relative contribution of the flat-spectrum component.
The `red' component is the spectrum produced by a 1-Gyr `Burst' model of
galaxy evolution at an age of 10 Gyr. The blue component is a power-law
with spectral index $\alpha$ = 0.2 ($f_{\nu} \propto \nu^{-\alpha}$),
the mean optical spectral index found for quasars by Barvainis (1990).
2355$-$010 is a red radio galaxy with only a very small value of $f_{5000}$,
while 0059$+$027 is one of the bluer galaxies in the PSR sample.}
{7.0}

This procedure is illustrated in Figure 6.
For the `old' or `red' component we chose to adopt a spectrum capable of
producing the reddest colours seen in radio galaxies at $z \simeq 1$
({\it e.g.} 3C65); in practice this was achieved using the spectrum produced
by a stellar population of age 10 Gyr in an updated version
of the models of Guiderdoni \& Rocca-Volmerange (1987).
For the `young' or `blue' component, we decided
to adopt a power-law spectrum ($f_{\nu} \propto \nu^{-\alpha}$)
with a spectral index $\alpha = 0.2$. This choice of spectrum can be justified
at two different levels. First, the exact value of
$\alpha$ was chosen in the spirit of scattered quasar light;
Barvainis (1990)
concluded that the mean value for the optical spectral index
in high luminosity quasars ({\it i.e.} those whose optical spectra
are essentially uncontaminated by a host galaxy contribution) is
$\alpha = 0.2$. Second, empirically,
this form of spectrum is an excellent
representation of the approximately
flat $f_{\nu}$ optical-UV continuum actually observed in high-redshift
radio galaxies.

\japfig{7}{Radio power, $P_{\rm 2.7\;GHz}$,
versus $f_{5000}$ for the combined
46-source 3CR/1-Jy/PSR sample. 3CR sources are shown as squares, 1-Jy sources
as triangles, and the PSR sources as circles. Notice that the
correlation is mainly in the sense of setting an upper limit to
$f_{5000}$ at given power.}
{9.0}

In Figure 7 we show the quantitative relation between this
definition of UV activity and radio power.
Radio power and $f_{5000}$ appear to be strongly correlated (no PSR galaxy
has $f_{5000} > 0.19$ whereas more than half the 3CR galaxies have
$f_{5000} > 0.20$).
This result contrasts sharply with that of Lilly (1989),
who reported that in his combined 3CR and 1-Jy sample there was no
significant correlation between $f_{5000}$ and $P_{\rm 408\;MHz}$.
The origin of the difference appears to be an
error in Lilly's calculation of radio luminosity.
An interesting aspect of the relation with power is
that all sub-samples appear to possess a
range of $f_{5000}$ values, but with power apparently setting the
upper limit in $f_{5000}$. This suggests the existence of
a second parameter which determines the actual level of
UV light -- see Dunlop \& Peacock (1993) for further
discussion.

For the present, the point to emphasise is that this diagram
provides a quantitative definition of a radio-quiet
galaxy. At least at $z\simeq 1$, any galaxy with
$P_{2.7}\ls 10^{25.5}\;\rm WHz^{-1}sr^{-1}$ (for $h=1/2$)
has a negligibly small level of UV activity. There
have been some suggestions that UV activity and alignments
are functions specifically of redshift, but there is
little evidence that this is anything other than a
reflection of the above trend in a flux-limited sample.
Until proven otherwise, the natural null hypothesis is
that galaxies below this power level at higher
redshifts also reflect the properties of the
general population of massive ellipticals.

\japfig{8}{Comparison of the
$R-K$ colours of the PSR galaxies (solid squares and
triangles) and the 3CR galaxies in the subsample (open circles and
diamonds).
PSR galaxies with measured redshifts are denoted by solid
triangles, those with estimated redshifts by solid squares.
3CR galaxies whose $K$-band morphologies are
aligned with $15^{\circ}$ of the radio axis are denoted by diamonds, and the
remainder by open circles.
Also shown are five 1-Jy galaxies (from Lilly 1989) (asterisks), and all
spectroscopically confirmed quasars with $0.5 < z < 2.0$ in the PSR sample
(stars).
The dashed line shows the effect of simply k-correcting the spectrum.
The solid line shows
a very old ($z_f = 50$, $\Omega_0 = 0$, $H_0 = 50\, {\rm km s^{-1}Mpc^{-1}}$)
UV-hot model of elliptical galaxy
evolution (Rocca-Volmerange 1989).}
{9.0}

In Figure 8 we compare the $R-K$ colours of the PSR and 3CR galaxies.
Several other objects
which are not part of our PSR and 3CR subsamples have been included
here for comparison purposes.
These are (i) the very red 3CR galaxy 3C65, (ii) the five 1-Jy galaxies with
measured redshifts for which $r-K$ colours are given by Lilly (1989),
and (iii) all spectroscopically confirmed quasars with $0.5 < z < 2.0$
in the Parkes Selected Regions sample for which $R-K$ colours exist
(Dunlop {\it et al.} 1989a).

This diagram displays a number of important features.
First, with the obvious exception
of 3C65, the PSR galaxies are consistently redder than the 3CR galaxies;
moreover, the PSR galaxies display
remarkably little dispersion in their optical-infrared colours.
This is well consistent with the findings of Rixon, Wall \& Benn
(1991) at lower redshift: they found the rest-frame
colours of radio ellipticals at $z<0.3$ to be constant
to within a few hundredths of a magnitude.
In contrast, the 3CR galaxies scatter downwards from the well-defined
PSR locus towards the region of colour space occupied by the PSR
quasars (the very red galaxies 3C65 and 1129$+$37 appear
to be exceptional).
Of the six 3CR galaxies with $R-K \leq 4.0$, all but one (3C252)
have $K$-band morphologies clearly aligned with their radio axes.

The homogeneity of the PSR galaxies, along with the lack of any
dramatic alignment effect in the redder galaxies, suggests instead that the
true
optical-infrared colour of a radio-quiet elliptical at $z \simeq 1$
is actually $R-K \simeq 4.8$. Values of
$f_{5000} \simeq 0.05$ might be a feature of most elliptical galaxies
at $z \simeq 1$.
This is certainly consistent with the results of Arag\'on-Salamanca
{\it et al.} (1993). From optical/IR photometry of clusters
of galaxies up to $z=0.8$, they conclude that ellipticals
(mainly radio-quiet) in the highest-redshift clusters
are slightly bluer than present-day ellipticals.
On the assumption that these galaxies formed in
a single burst, their data allow the epoch of formation
to be as low as $z=2$. However, the radio-selected
samples extend the range still further. Although the above discussion
has concentrated on the situation at $z\simeq 1$, the PSR
sample contains a number of galaxies inferred from
colour-estimated redshifts and from the $K$--$z$ relation to
have $z\simeq 2$. These also are apparently old and red,
with $R-K\simeq 4$ -- 5. If this is taken to imply a minimum
age of 1$h^{-1}$ Gyr, the formation redshift is pushed out to
between 3.3 and 7.2, depending on $\Omega$.
Note that this is the epoch at which the whole galaxy
must be assembled: ellipticals cannot have been
assembled from many small clumps after star formation
had ceased (Bower, Lucey \& Ellis 1992).
It will be
fascinating to pursue this line of argument in mJy samples,
where we may hope to find `normal' radio galaxies at
$z>3$. If these are still red, the consequences for
galaxy formation models will be radical indeed.

\sec{7. CONCLUSIONS}

This review has given a brief summary of the properties
of galaxies as viewed in the radio background. In conclusion,
it is worth emphasising three points:
\item{(i)} Although some factors such as galaxy mass
and Hubble type strongly dispose a galaxy to host
a radio-loud AGN, we still have no definite understanding
of why this should be so. Other `distinguishing marks'
of radio galaxies might be helpful in this process,
but few if any are clearly established.
\item{(ii)} With certain exceptions (such as the situation
at $z\gs 2$), we have a good statistical description
of how the abundance of radio AGN evolves. Again, though,
we are very far from understanding why active nuclei
found it so much easier to function at high redshift.
\item{(iii)} High-redshift radio galaxies should probably not
be thought of as in any way primaeval. If we ignore the
few dozen most luminous sources in the universe, then
the optical/IR properties of high-redshift radio
galaxies are consistent with those of radio-quiet
ellipticals. They appear to be red and old: theories
in which most massive galaxies complete their star
formation at $z\gs 4$ are required.

\sec{REFERENCES}

\japref Arag\'on-Salamanca, A., Ellis, R.S. \& Sharples, R.M., 1991.  {\it Mon.
Not. R. astr. Soc.}, {\bf 248}, 128.
\japref Arag\'on-Salamanca, A., Ellis, R.S. Couch, W.J. \& Carter, D., 1993.
{\it Mon. Not. R. astr. Soc.}, {\bf 262}, 764.
\japref Barvainis, R., 1990. {\it Astrophys. J.}, {\bf 353}, 419.
\japref Bebbington, D.H.O., 1986.  {\it Mon. Not. R. astr. Soc.}, {\bf 218},
577.
\japref Blandford, R.D., 1990. {\it Active galactic nuclei}, 20th SAAS-FEE
lectures.
\japref Bower, R.G., Lucey, J.R. \& Ellis, R.S., 1992.  {\it Mon. Not. R. astr.
Soc.}, {\bf 254}, 601.
\japref Boyle, B.J., Shanks, T. \& Peterson, B.A., 1987.  {\it Mon. Not. R.
astr. Soc.}, {\bf 235}, 935.
\japref Broadhurst, T.J., Ellis, R.S. \& Glazebrook, K., 1992. {\it Nature},
{\bf 355}, 55.
\japref Carlberg, R.G., 1990. {\it Astrophys. J.}, {\bf 350}, 505.
\japref Coles, P., 1992. {\it Comments on Astrophys.}, {\bf 16}, 45.
\japref Chambers, K.C. \& Charlot, S., 1990. {\it Astrophys. J.}, {\bf 348},
L1.
\japref Chambers, K.C., Miley, G. \& Joyce, R., 1988. {\it Astrophys. J.}, {\bf
329}, L75.
\japref Chambers, K.C., Miley, G.K. \& van Breugel, W.J.M., 1987, {\it Nature},
{\bf 329}, 624.
\japref Cimatti, A., di Serego Alighieri, S., Fosbury, R.A.E., Salvati, M. \&
Taylor, D., 1993.  {\it Mon. Not. R. astr. Soc.}, in press.
\japref Daly, R.A., 1992. {\it Astrophys. J.}, {\bf 399}, 426.
\japref Davies, R.D., Pedlar, A. \& Mirabel, I.F., 1978.  {\it Mon. Not. R.
astr. Soc.}, {\bf 182}, 727.
\japref Disney, M.J. \& Sparks, W.B., 1984. {\it Mon. Not. R. astr. Soc.}, {\bf
206}, 899.
\japref Dunlop, J.S., Peacock, J.A., Savage, A., Lilly, S.J., Heasley, J.N. \&
Simon, A.J.B., 1989a. {\it Mon. Not. R. astr. Soc.}, {\bf 238}, 1171.
\japref Dunlop, J.S., Guiderdoni, B., Rocca-Volmerange, B., Peacock, J.A. \&
Longair, M.S., 1989b. {\it Mon. Not. R. astr. Soc.}, {\bf 240}, 257.
\japref Dunlop, J.S. \& Peacock, J.A., 1990. {\it Mon. Not. R. astr. Soc.},
{\bf247}, 19.
\japref Dunlop, J.S. \& Peacock, J.A., 1993. {\it Mon. Not. R. astr. Soc.}, in
press.
\japref Eales, S.A. \& Rawlings, S., 1990. {\it Mon. Not. R. astr. Soc.}, {\bf
243}, 1P.
\japref Eales, S., Rawlings, S., Puxley, P., Rocca-Volmerange, B. \& Kuntz, K.,
1993. {\it Nature}, {\bf 363}, 140.
\japref Efstathiou, G. \& Rees, M.J., 1988. {\it Mon. Not. R. astr. Soc.}, {\bf
230}, 5P.
\japref Eisenhardt, P. \& Chokshi, A., 1990. {\it Astrophys. J.}, {\bf 351},
L9.
\japref Fanaroff, B.L. \& Riley, J.M., 1974. {\it Mon. Not. R. astr. Soc.},
{\bf 167}, 318.
\japref Glazebrook, K., 1991. PhD thesis, Univ. of Edinburgh.
\japref Goldschmidt, P., Miller, L.,  La Franca, F. \& Cristiani, S., 1992.
{\it Mon. Not. R. astr. Soc.}, {\bf 256}, 65P.
\japref Guiderdoni, B. \& Rocca-Volmerange, B., 1987. {\it Astr. Astrophys.},
{\bf 186}, 1.
\japref Heckman, T.M., Illingworth, G.D., Miley, G.K. \& van Breugel, W.J.M.,
1985. {\it Astrophys. J.}, {\bf 299}, 41.
\japref Heckman, T.M., Chambers, K.C. \& Postman, M., 1992. {\it Astrophys.
J.}, {\bf 391}, 39.
\japref Jannuzi, B.T. \& Elston, R., 1991. {\it Astrophys. J.}, {\bf 366}, L69.
\japref Jenkins, C.R., 1984. {\it Mon. Not. R. astr. Soc.},  {\bf 207}, 445.
\japref Lacey, C. \& Cole, S., 1993. {\it Mon. Not. R. astr. Soc.}, {\bf 262},
627.
\japref Lauer, T.R. {\it et al.}, 1992. {\it Astr. J.}, {\bf 103}, 703.
\japref Lilly, S.J. \& Longair, M.S., 1984. {\it Mon. Not. R. astr. Soc.}, {\bf
211}, 833.
\japref Lilly, S.J., 1988. {\it Astrophys. J.}, {\bf 333}, 161.
\japref Lilly, S.J., 1989. {\it Astrophys. J.}, {\bf 340}, 77.
\japref Longair, M.S., 1966. {\it Mon. Not. R. astr. Soc.}, {\bf 133}, 421.
\japref Longair, M.S., 1978. {\it Observational cosmology}, 8th SAAS-FEE
lectures.
\japref Loveday, J., Peterson, B.A., Efstathiou, G.P. \& Maddox, S.J., 1992.
{\it Astrophys. J.}, {\bf 390}, 338.
\japref McCarthy, P.J., van Breugel, W., Spinrad, H. \& Djorgovski, S., 1987.
{\it Astrophys. J.}, {\bf 321}, L29.
\japref Owen, F.N. \& White, R.A., 1991. {\it Mon. Not. R. astr. Soc.}, {\bf
249}, 164.
\japref Peacock, J.A., 1991. {\it Mon. Not. R. astr. Soc.}, {\bf 253}, 1P.
\japref Press, W.H. \& Schechter, P., 1974. {\it Astrophys. J.}, {\bf 187},
425.
\japref Prestage, R.M. \& Peacock, J.A., 1989. {\it Mon. Not. R. astr. Soc.},
{\bf 230}, 131.
\japref Rigler, M.A., Lilly, S.J., Stockton, A., Hammer, F. \& Le F\`evre O.,
1992.  {\it Astrophys. J.}, {\it 385}, 61.
\japref Rixon, G.T., Wall, J.V. \& Benn, C.R., 1991. {\it Mon. Not. R. astr.
Soc.},  {\bf 251}, 243.
\japref Rocca-Volmerange, B., 1989. {\it Mon. Not. R. astr. Soc.}, {\bf 236},
47.
\japref Romanishin, W. \& Hintzin, P., 1989. {\it Astrophys. J.}, {\bf 341},
41.
\japref Rowan-Robinson, M., Benn, C.R., Broadhurst, T.J., Lawrence, A. \&
McMahon, R.G., 1993. {\it Mon. Not. R. astr. Soc.}, {\bf 263}, 123.
\japref Sadler, E.M., Jenkins, C.R. \& Kotanyi, C.G., 1989. {\it Mon. Not. R.
astr. Soc.}, {\bf 240}, 591.
\japref Sansom, A., Wall, J.V. \& Sparks, W.B., 1987. {\it Structure \&
dynamics of elliptical galaxies}, IAU Symp. no 127, ed T. de Zeeuw (D. Reidel),
p429.
\japref Saunders, W., Rowan-Robinson, M., Lawrence, A., Efstathiou, G.,
Kaiser, N., Ellis, R.S. \& Frenk, C.S., 1990. {\it Mon. Not. R. astr. Soc.},
{\bf 242}, 318.
\japref Smith, E.P. \& Heckman, T.M., 1989. {\it Astrophys. J.}, {\bf 341},
658.
\japref Smith, E.P. \& Heckman, T.M., 1990. {\it Astrophys. J.}, {\bf 348}, 38.
\japref Smith, E.P., Heckman, T.M. \& Illingworth, G.D., 1990. {\it Astrophys.
J.}, {\bf 356}, 399.
\japref Soltan, A., 1982. {\it Mon. Not. R. astr. Soc.},  {\bf 200}, 115.
\japref Sparks, W.B., 1984. {\it Mon. Not. R. astr. Soc.},  {\bf 204}, 1049.
\japref Sparks, W.B., Disney, M.J., Wall, J.V. \& Rodgers, A.W., 1984. {\it
Mon. Not. R. astr. Soc.},  {\bf 207}, 445.
\japref Spinrad, H. \& Djorgovski, S., 1987. in Hewitt A., Burbidge G., Fang
L.Z.,
eds, Proc. IAU Symp. 124, Observational Cosmology. Reidel,
Dordrecht, p129.
\japref Subramanian, K. \& Padmanabhan, T., 1993. {\it Mon. Not. R. astr.
Soc.}, in press.
\japref Tadhunter, C.N., Scarrott, S.M., Draper, P. \& Rolph C., 1992. {\it
Mon. Not. R. astr. Soc.}, { \bf 256}, 53P.
\japref Uson, J.M., Bagri, D.S., \& Cornwell, T. J., 1991. {\it Phys. Rev.
Lett.}, {\bf 67}, 3328.
\japref Wall, J.V., Pearson, T.J. \& Longair, M.S., 1980. {\it Mon. Not. R.
astr. Soc.}, {\bf 193}, 683.
\japref Wieringa, M.H., de Bruyn, A.G. \& Katgert, P., 1992. {\it Astr.
Astrophys.}, {\bf 256}, 331.
\japref Windhorst, R., 1984. PhD thesis, Univ. of Leiden.
\japref Windhorst, R.A., Koo, D.C., Spinrad, H., 1986. in
Madore B.F., Tully R.B., eds, NATO \& Advanced Research Workshop,
Galaxy distances and Deviation from Universal Expansion.
Reidel, Dordrecht, p197.
\japref Wolfe, A.M., 1993. {\it Astrophys. J.}, {\bf 402}, 411.
\bye